\documentclass[prd,twocolumn,nofootinbib]{revtex4}
\usepackage{graphicx, epsfig}
\usepackage{color}
\usepackage{mathrsfs}
\usepackage {amssymb}
\usepackage {amsmath}
\usepackage{soul}

\newcommand{\nc}{\newcommand}
\newcommand{\nn}{\nonumber}

\nc{\half}{\frac{1}{2}}
\def\M{M_{\rm Pl}}
\def\Psc{P_{{\rm s}}}
\def\Ito{It\^o\,\,}
\def\I{{\rm I}}
\def\S{{\rm S}}
\newcommand{\refeq}[1]{(\ref{#1})}

\usepackage{bm}
\usepackage{braket}

\newcommand{\dif}[2]{\frac{\mathrm{d} #1}{\mathrm{d} #2}}
\newcommand{\pdif}[2]{\frac{\partial #1}{\partial #2}}
\newcommand{\Mpl}{M_\text{Pl}}
\newcommand{\FP}{\text{FP}}
\newcommand{\dd}{\mathrm{d}}
\newcommand{\cl}{\text{cl}}
\newcommand{\ns}{n_{{}_\mathrm{S}}}

\newcommand{\calC}{\mathcal{C}}
\newcommand{\uf}{\text{f}}
\newcommand{\calL}{\mathcal{L}}
\newcommand{\calM}{\mathcal{M}}
\newcommand{\calN}{\mathcal{N}}
\newcommand{\calP}{\mathcal{P}}

\newcommand{\bae}[1]{\begin{align} #1 \end{align}}
\newcommand{\bce}[1]{\begin{cases} #1 \end{cases}}
\newcommand{\dps}{\displaystyle}
\newcommand{\bpme}[1]{\begin{pmatrix} #1 \end{pmatrix}}

\input{epsf.sty}

\begin{document}

\title{Inflationary stochastic anomalies}

\author{Lucas Pinol$^{1,2}$, S\'ebastien Renaux-Petel$^{1}$ and Yuichiro Tada$^{1,3}$}

\affiliation{$^1$Institut d'Astrophysique de Paris, UMR 7095 du CNRS et Sorbonne Universit\'e, \\
98~bis~bd~Arago, 75014 Paris, France}
\affiliation{$^2$ \'Ecole Normale Sup\' erieure, 45 rue d'Ulm, 75005 Paris, France}
\affiliation{$^3$Department of Physics, Nagoya University, Nagoya 464-8602, Japan}

\date{\today}

\begin{abstract} The stochastic approach aims at describing the long-wavelength part of quantum fields during inflation by a classical stochastic theory. It is usually formulated in terms of Langevin equations, giving rise to a Fokker-Planck equation for the probability distribution function of the fields, and possibly their momenta. The link between these two descriptions is ambiguous in general, as it depends on an implicit discretisation procedure, the two prominent ones being the \Ito and Stratonovich prescriptions. Here we show that the requirement of general covariance under field redefinitions is verified only in the latter case, however at the expense of introducing spurious `frame' dependences. This stochastic anomaly disappears when there is only one source of stochasticity, like in slow-roll single-field inflation, but manifests itself when taking into account the full phase space, or in the presence of multiple fields. Despite these difficulties, we use physical arguments to write down a covariant Fokker-Planck equation that describes the diffusion of light scalar fields in non-linear sigma models in the overdamped limit. 
We apply it to test scalar fields in de Sitter space and show that some statistical properties of a class of two-field models with derivative interactions can be reproduced by using a correspondence with a single-field model endowed with an effective potential. We also present explicit results in a simple extension of the single-field $\lambda \phi^4$ theory to a hyperbolic field space geometry.
The difficulties we describe seem to be the stochastic
counterparts of the notoriously difficult problem of maintaining general covariance in quantum
theories, and the related choices of operator ordering and path-integral constructions. Our work thus
opens new avenues of research at the crossroad between cosmology, statistical physics,
and quantum field theory.
\end{abstract}

\maketitle


{\bf Introduction.}--- The theory of cosmological inflation can boast from a spectacular success to explain the patterns observed in the Cosmic Microwave Background anisotropies \cite{Ade:2015lrj,Ade:2015ava}. According to it, cosmological fluctuations originate from microscopic quantum fluctuations stretched to cosmological scales. This paradigm is therefore deeply rooted both in quantum field theory (QFT) and in general relativity. However, in its conventional formulation, the geometry of spacetime and the fields active during inflation are split between an homogeneous background, obeying classical equations of motion, and spatial fluctuations, which are treated quantum-mechanically \cite{Weinberg:2008zzc,Peter:1208401}.
In the absence of a full theory of quantum gravity, this practical approximation scheme is justified in situations in which the classical behaviour dominates the dynamics. Conceptually, it is not entirely satisfactory though, and moreover, it is expected to break down in the presence of very light scalar fields. 
The stochastic formalism aims at addressing this issue by deriving a classical effective theory for the coarse-grained super-Hubble part of the quantum fields driving inflation \cite{Starobinsky:1986fx,Starobinsky:1994bd}. The expansion of the universe results in a continuous flow of initially sub-Hubble modes joining the super-Hubble sector, which gives rises  in this framework to a stochastic dynamics. Stochastic inflation is at the heart of the concepts of eternal chaotic inflation and the multiverse \cite{Linde:1986fc,Linde:1986fd,Goncharov:1987ir}, with possibly far-reaching consequences. Moreover, as the production of primordial black holes during inflation necessitates very flat potentials, in which quantum diffusion effects are non-negligible, it is also necessary to take it into account in the context of the identification of the nature and origin of dark matter \cite{Kawasaki:2015ppx,Pattison:2017mbe,Biagetti:2018pjj,Ezquiaga:2018gbw}. Scrutinising the theoretical grounds of stochastic inflation is therefore conceptually and practically very important, and it has recently received a renewed attention \cite{Burgess:2014eoa,Vennin:2015hra,Burgess:2015ajz,Vennin:2016wnk,Moss:2016uix,Hardwick:2017fjo,Grain:2017dqa,Collins:2017haz,Prokopec:2017vxx,Tokuda:2017fdh,Hardwick:2018sck,Tokuda:2018eqs}. In this Letter, we summarize the salient features of a detailed investigation \cite{prep1,prep2}, highlighting in particular the hitherto unnoticed conceptual issue of formulating a generally covariant theory of stochastic inflation. This difficulty is related to the \Ito versus Stratonovich dilemma in statistical physics \cite{dilemma}, and ultimately to the notoriously difficult problem of studying path integral and quantum anomalies in curved target spaces \cite{1957RvMP...29..377D,AlvarezGaume:1983ig,Grosche-87,Bastianelli:2006rx}.\\


{\bf \Ito versus Stratonovich.}--- It is well known that stochastic differential equations (SDE) are not completely defined in general unless they are accompanied with a prescription to make sense of their stochasticity. Different types of consistent stochastic calculus have been defined by mathematicians, chief amongst them \Ito and Stratonovich calculus \cite{Ito,Strato}, that physically correspond to different discretisation procedures for stochastic processes\footnote{We concentrate on these two well known discretisations, but one can easily consider more general $\alpha$-discretisatons \cite{prep2}, where $0 \leq \alpha \leq 1$, recovering \Ito calculus for $\alpha=0$ and the Stratonovich one for $\alpha=1/2$.} (see \textit{e.g.} \cite{risken1989fpe,vankampen2007spp}). It will prove useful to present them in a rather general way. For this, let us consider an arbitrary number of random variables $X^a$, which depend on time $N$, and that obey the generic Langevin equations
\begin{equation}
\frac{{\rm d} X^a}{{\rm d} N}=h^a  + g^a_\alpha \, \xi^\alpha\,,
\label{Langevin}
\end{equation}
where $\xi^\alpha$ are independent normalised Gaussian white noises, verifying $\langle \xi^\alpha(N) \xi^\beta(N') \rangle=\delta^{\alpha \beta} \delta(N-N')$, and whose numbers need not be the same as the number of $X^a$'s. Both the deterministic and stochastic parts $h^a$ and $g^a_\alpha$ are in general functions of the fields $X^a$ and of $N$, although we will omit to mention the time-dependence for simplicity. In a discretised version, the physical ambiguity comes from whether the strengths of the random kicks are determined by the amplitudes of the noises $g^a_\alpha$ at a time immediately before the kicks --- this is the \Ito choice --- or, motivated by the fact that white noises are idealisations of random processes with finite correlation time, whether they are determined by the average of the noises amplitudes over the duration of the kick --- this is the Stratonovich choice.
The two procedures are physically distinct, and the corresponding probability distribution functions $P(\bm{X},N)$ (PDF) verify the distinct Fokker-Planck (FP) equation
\begin{eqnarray}
\frac{\partial P}{\partial N}&=&{\cal L}_{{\rm FP}}(\bm{X}) \cdot P \nn \\ 
{\rm with} \quad {\cal L}_{{\rm FP}}(\bm{X})&=&-\frac{\partial}{\partial X^a} D^a+\frac12 \frac{\partial^2}{\partial X^a \partial X^b} D^{ab}\,,
\label{FP}
\end{eqnarray}
where the so-called drift $D^a$ is given by $D^a_\I=h^a$ in the \Ito prescription, and by $D^a_\S=h^a+\frac12 g^b_\alpha \partial g^a_\alpha /\partial X^b$ in the Stratonovich one, and where the diffusion coefficient reads $D^{ab}=g^a_\alpha g^b_\alpha$ in both cases. The two frameworks differ in the presence of multiplicative noises, \textit{i.e.} when the $g^a_\alpha$ depend on the random variables $X^a$, which results in the noise-induced drift in the Stratonovich prescription, in addition to the deterministic drift. The Stratonovich FP operator can be rewritten as ${\cal L}_{{\rm FP, S}} \cdot P =-\partial_a( h^a P) +\frac12\partial_a \left(  g^a_\alpha \partial_b (g^b_\alpha P) \right)$, so that the choice between the two prescriptions is reminiscent of the factor ordering problem in QFT. Mathematically, a given set of SDE can always be rewritten in either of the two prescriptions by changing the expression of $h^a$ in the Langevin equations \eqref{Langevin}, but physically the deterministic part of the evolution is often specified, so that other physical considerations should be taken into account to fix the ambiguity. 

In the context of inflation, it has been argued that the \Ito prescription should be preferred to respect causality \cite{Fujita:2014tja,Tokuda:2017fdh}, although the Stratonovich choice is in no way problematic in this respect. On the other hand, the Stratonovich prescription has been advocated by the fact that the white noises should be treated as a limit of colored noises when the smooth decomposition between short and long-wavelength modes becomes sharp \cite{Mezhlumian:1991hw}.
It has been also argued that the choice between the two prescriptions exceeds the accuracy of the stochastic approach \cite{discussion-Starobinsky,Vennin:2015hra}. We will show that while this latter point is indeed true in a certain sense for slow-roll single-field inflation, the \Ito versus Stratonovich dilemma strikes back in a non-trivial manner as soon as more than one degree of freedom is involved, in the full phase space or when multiple fields are taken into account.\\


{\bf Slow-roll single-field inflation.}--- We begin by considering the simple framework of slow-roll single-field inflation, with the corresponding Langevin equation \cite{Starobinsky:1986fx}
\begin{eqnarray}
\frac{{\rm d} \phi}{{\rm d} N}=-\frac{V_{,\phi}}{3 H^2( \phi)}+\frac{H(\phi)}{2 \pi} \xi
\label{Langevin-single-field}
\end{eqnarray}
where $3 H^2(\phi) \M^2=V(\phi)$. Here, $\phi$ is the coarse-grained scalar field, $V(\phi)$ denotes its potential, the deterministic drift describes the overdamped, slow-roll, behaviour, and the amplitude of the noise term is given by the size of super-Hubble fluctuations of a light scalar field (we will discuss below the origin and regime of validity of Eq.~\eqref{Langevin-single-field} and consider more general situations). It is now well understood that the number of $e$-folds of expansion $N \propto \ln a$, where $a$ is the scale factor, should be used as the time variable in stochastic inflation, in order to agree with cosmological perturbation theory and to warrant the agreement with perturbative quantum field theory \cite{Finelli:2008zg,Finelli:2010sh,Vennin:2015hra}. 
Note that there is a considerable conceptual and technical difference between test and non-test scalar fields. If a test scalar field in a fixed spacetime geometry is considered, 
the instances of $H(\phi)$ in \eqref{Langevin-single-field} should be replaced by the deterministically determined Hubble scale $H(N)$, which is constant in the most studied case of de Sitter spacetime. 
The stochastic formalism can then be seen as a powerful method to resum infrared divergences of a light quantum scalar field in de Sitter spacetime.
In addition, in the case of a scalar field driving inflation, the Langevin equation \eqref{Langevin-single-field} has been used to compute various observable quantities like the power spectrum of the curvature perturbation, using the techniques of first passage time analysis, recovering results of conventional cosmological perturbation theory in a suitable classical limit \cite{Vennin:2015hra}.
Using the same techniques, we show that the \Ito or Stratonovich prescriptions lead in this context to the same results to an excellent accuracy. 
Let ${\cal N}(\phi)$ be the number of $e$-folds of inflation that is realised by starting from the initial value $\phi$, from which the distribution of primordial density perturbations can be easily computed with the stochastic-$\delta N$ formalism \cite{Fujita:2013cna,Fujita:2014tja,Vennin:2015hra,Assadullahi:2016gkk}. The PDF of this stochastic variable is equivalently defined by its Fourier transform, the characteristic function $\chi_{{\cal N}}(\omega,\phi) \equiv \langle e^{i \omega {\cal N}(\phi)} \rangle$.
From the FP equation,
one can show that $\chi_{\cal N}$ satisfies ${\cal L}^\dagger_{{\rm FP}}(\phi)\chi_{{\cal N}}(\omega,\phi)=- i \omega\, \chi_{{\cal N}}(\omega,\phi)$, where $\dagger$ indicates the adjoint operator. Whereas 
${\cal L}^\dagger_{{\rm FP, I}}/\M^2=v\frac{\partial^2}{\partial \phi^2}-\frac{v'}{v} \frac{ \partial }{ \partial \phi}$, one has ${\cal L}^\dagger_{{\rm FP, S}}/\M^2=v\frac{\partial^2}{\partial \phi^2}-\frac{v'}{v}\left(1-\frac{v}{2}\right) \frac{ \partial }{ \partial \phi}$, with the same boundary conditions, and where $v \equiv V/(24 \pi^2 \M^4)$ is the dimensionless potential in Planck units. As $v \ll 1$ for consistently working in the classical gravity regime, one can see that in the regime of validity of stochastic inflation, the \Ito or Stratonovich prescriptions for slow-roll single-field inflation lead to the same results, both when the classical result of standard perturbation theory is recovered, and when stochastic effects are large. This can be confirmed by deriving recursive differential equations between the moments of ${\cal N}$, and computing numerically all related observables like the power spectrum and the scalar spectral index (see appendix \ref{details-single-field}).\\


{\bf General covariance.}--- The physical criterion that we use to shed new light on the conceptual issues of stochastic inflation is the one of its covariance under general coordinate transformations. The conundrum can be stated rather concisely: on one hand, \Ito calculus does not verify the standard chain rule, and the corresponding Fokker-Planck equation does not respect general covariance under field redefinitions. On the other hand, Stratonovich calculus does respect general covariance, but at the expense of introducing spurious `frame' dependences.\footnote{\Ito calculus has nothing particular in this respect, as only Stratonovich calculus respects general covariance in the general class of $\alpha$-discretisations.}
In both cases, a classical symmetry is broken by the inclusion of stochastic effects, a phenomenon we refer to as a stochastic anomaly.

To present the issue in a simple but rather general form, let us consider the generalisation of \refeq{Langevin-single-field} to non-linear sigma models (NLSM),
with Lagrangian ${\cal L}=-\frac12 G_{IJ} \partial_\mu \phi^I \partial^\mu \phi^J-V$, assuming for simplicity that all properly normalised fields are effectively light. If one uses the slow-roll classical equation of motion to determine the drift, such a generalisation reads
\begin{equation}
\frac{{\rm d} \phi^I}{{\rm d} N}=-\frac{G^{IJ} V_{,J}}{3 H^2(\boldsymbol{\phi})}+\Xi^I\,,
\label{Langevin-multi-field}
\end{equation}
where $\Xi^I$ are stochastic noises, $3 H^2(\boldsymbol{\phi}) \M^2=V(\boldsymbol{\phi})$, and $\boldsymbol{\phi}$ denotes the set of fields. 
A crucial point is that any derivation of this equation or generalisations  thereof --- be it from a heuristic approach at the level of the equations of motion, or from the more elaborate perspective of deriving a coarse-grained action within the Schwinger-Keldysh formalism \cite{prep1} --- does not result in Langevin equations of the form \eqref{Langevin}, but only determines the correlation functions of the noises, \textit{i.e.} determines the diffusion matrix $D^{ab}$, and not a set of `square-roots' $g^{a}_\alpha$, which can differ by arbitrary field-dependent rotations $g^{a}_\alpha \to \Omega_\alpha^{\,\beta}(\boldsymbol{X}) g^{a}_\beta$.
For instance, in a generalised slow-roll approximation, one obtains
\begin{eqnarray}
\langle \Xi^I(N)\, \Xi^J(N') \rangle=\left(\frac{H}{2 \pi} \right)^2 G^{IJ} \,\delta(N-N')\,,
\label{slow-roll-noise-correlations}
\end{eqnarray}
but no prescription to write down $\Xi^I$ as a weighted sum of independent white noises. The correlations \eqref{slow-roll-noise-correlations} can be realised by writing $\Xi^I=g^I_\alpha \, \xi^\alpha$ with $g^I_\alpha=\frac{H}{2 \pi}e^I_\alpha$, for \textit{any} set of vielbeins of the field space metric. This arbitrariness is innocuous in \Ito calculus, as the precise expression of $g^a_\alpha$ does not alter the FP equation \eqref{FP}, which depends only on the deterministic drift and on the diffusion matrix.
It does matter in the Stratonovich interpretation however, as the field space derivatives of $g^a_\alpha$ enter into the FP equation. 

Now comes the issue of general covariance: physical quantities should not depend on arbitrary field redefinitions that one can perform at the level of the action. For NLSM studied here, the fields $\phi^I$ are indeed merely coordinates on the field space manifold of metric $G_{IJ}$. To discuss the covariance of the FP equation, it is convenient to express it in terms of the rescaled PDF $\Psc=P(\boldsymbol{\phi})/\sqrt{{\rm det} (G_{IJ})}$, which is a scalar under general field redefinitions $\boldsymbol{\phi} \to \boldsymbol{\tilde{\phi}}(\boldsymbol{\phi})$, 
and in terms of which any sensible stochastic theory should be manifestly covariant. Using the above equations, one obtains that this required property is not satisfied in the \Ito interpretation. 
One way to understand this is to note that in \Ito calculus, the drift $h^a$ in the Langevin equation \eqref{Langevin} does not transform as a vector under redefinitions $X^a \to \tilde{X}^a$, whereas the drift in Eq.~\eqref{Langevin-multi-field} does classically transform as a vector under field space redefinitions (see \cite{prep2} for details).
If one uses the Stratonovich interpretation however, one obtains the manifestly covariant result
\begin{eqnarray}
\frac{\partial \Psc}{\partial N}&=& \nabla_I \left( \frac{V^{,I}}{3 H^2} \Psc \right)+\frac{1}{2} \nabla_I \left( \frac{H}{2 \pi}  \nabla^I \left( \frac{H}{2 \pi}  \Psc \right) \right) \nn \\
&+&\frac{1}{2} \nabla_I \left( \left(\frac{H}{2 \pi}\right)^2 e^I_\alpha (\nabla_J e^J_\alpha)\, \Psc \right)
\label{FP-Strato}
\end{eqnarray}
where field space indices are raised with the inverse metric, and $\nabla_I$ denotes the covariant derivative with respect to the field space metric. As announced, through their covariant derivatives in the second line, this comes however at the expense of a spurious dependence on the arbitrary choice of vielbeins, which can result in qualitatively very different results (see appendix \ref{frame-dependence}). Note also that for a general metric, no privileged frame exists that would make this effect disappear. 
More generally, had we considered a more general situation in which the right-hand side of Eq.~\eqref{slow-roll-noise-correlations} is different, this would not have changed the fact that the sub-Hubble physics dictates only $\langle \Xi^I(N)\, \Xi^J(N') \rangle$, and that the Stratonovich interpretation, necessary to maintain general covariance, forces us to introduce an artificial structure that does not drop out from physical observables without further modifications, be it in a curved or flat field space. 
It is also interesting to use this general point of view to shed different light on our discussion of single-field slow-roll inflation. If one allows field redefinition in that case, at the somewhat artificial expense of not having a standard kinetic term, different `coordinates' would yield different results in the \Ito interpretation, because of its lack of general covariance. Such a difficulty disappears in the Stratonovich one however, as in one dimension, there is no ambiguity in defining the square-root $g^a_\alpha$ up to an irrelevant sign. \\


{\bf First principle `derivation' in phase space.}--- An attempt to derive the stochastic theory from first principles in a general setup enables us to spell out various assumptions behind such a description, which is important in order to delineate its regime of validity, and shows that the difficulties described above hold true in more complete descriptions.
Leaving details to \cite{prep1}, using the Schwinger-Keldysh formalism, one derives there a coarse-grained action for the long-wavelength part of fields in NLSM, taking into account the full phase space using the Hamiltonian language, as well as the coupling to gravity, unifying and generalising previous works \cite{Morikawa:1989xz,Moss:2016uix,Grain:2017dqa,Tokuda:2017fdh,Tokuda:2018eqs}. Assuming that an effective classical description emerges, this gives rises to the following set of equations:
\begin{eqnarray}
\phi^{I \prime}=\frac{G^{IJ} \pi_J}{H} + \xi_Q^I \,, \quad D_N \pi_I=-3 \pi_I-\frac{V_I}{H}+\xi_{{\tilde P} I}
\end{eqnarray}
where $\pi_I$ represent the conjugate momentum of $\phi^I$ (divided by $a^3$), $'={\rm d}/{\rm d} N$, $D_N \pi_I=\pi_I^{\prime}-\Gamma_{IJ}^K \phi^{J \prime} \pi_K$, $3 H^2 \M^2=V(\boldsymbol{\phi})+\frac12 G^{IJ} \pi_I  \pi_J $, and where the two-point correlation functions of the stochastic noises $(\xi_Q^I,\xi_{{\tilde P} I})$ can be computed in principle \cite{prep1}. This system of equations is not of the type \eqref{Langevin}, for several reasons. 
The dynamics is strictly speaking non-Markovian \cite{Gautier:2012vh}, as the statistical properties of the noises do not depend only on the current location $(\phi^I,\pi_I)$ in phase space, but on the entire past history for each realisation. Indeed, the physics of the small scale fluctuations is dictated by the coarse-grained dynamics, which itself depends on all the stochastic kicks it has received.  
The stochastic noises are also in general non-Gaussian --- a feature that would give rise not to a FP but to a more general Kramers-Moyal equation \cite{risken1989fpe,Riotto:2011sf} --- and colored, owing to the smooth splitting between the long wavelength modes and the short ones that are integrated out \cite{Winitzki:1999ve,Matarrese:2003ye,Liguori:2004fa}. Even when the assumptions of a Markovian dynamics sourced by Gaussian white noises are adequate, the subtlety pointed out above is still present: one needs to introduce by hand `square-roots' of the correlation matrix of the noises, $g^I_{Q  \alpha}$ and $g_{{\tilde P} I \alpha}$, which transform as field-space vectors and covectors under field redefinitions \cite{prep2}. They do not appear in the \Ito FP equation, which however does not respect the criterion of general covariance, but explicitly appear in the covariant Stratonovich FP equation:
\begin{eqnarray}
\frac{\partial P}{\partial N}&=& -\frac{\pi^I}{H}D_{\phi^I} P+\left(\frac{V_{,I}}{H} +3 \pi_I \right) \partial_{\pi_I} P+3\, n P \nn \\
&+& \frac{1}{2}\left( D_{\phi^I} \left[ g^I_{Q  \alpha} \cdot \right]+ \partial_{\pi_I} \left[ g_{{\tilde P} I \alpha}  \cdot \right] \right) \nn \\
&&\left(D_{\phi^J} (g^J_{Q  \alpha} P)+\partial_{\pi_J}(g_{{\tilde P} I \alpha} P) \right)  \,,
 \label{FP-phase-space}
\end{eqnarray}
where the phase space PDF $P(\phi^I,\pi_I)$ is a proper scalar under field redefinitions, $n$ is the number of fields, and $D_{\phi^I}=\nabla_{I} +\Gamma^{K}_{I J} \pi_K \partial_{\pi_J}$ and $\partial_{\pi_I}$ are covariant derivatives in phase space. In addition to the arbitrariness that we point out in the diffusion term, note that the first line of \eqref{FP-phase-space} provides a neat reformulation of the deterministic evolution in non-linear sigma-models. It has a similar formal structure as a covariant Boltzmann equation in curved spacetime \cite{Straumann}, and can be useful beyond stochastic inflation, for instance to study
attractor behaviour and the sensitivity to initial conditions. In Ref.~\cite{prep2}, we also discuss the relationship between the full phase space description here and the overdamped limit in the previous section, and in particular how the latter can be approximately derived from the former in a suitable limit.\\


{\bf Stochastic diffusion in curved field space.}--- The various limitations of stochastic inflation that we have described should not diminish its successes, notably to reproduce non-trivial results from QFT in curved spacetime \cite{Tsamis:2005hd,Finelli:2008zg,Garbrecht:2013coa,Garbrecht:2014dca}. 
In the current lack of a more fundamental understanding, we assume that the covariant anomaly, manifested by the dependence on the arbitrary choice of vielbeins in \eqref{FP-Strato}, is a limitation of the stochastic description, and does not hold in the full quantum theory. Hence, we dismiss the last term in \eqref{FP-Strato}, which is technically equivalent to adding a suitable noise-induced term to the deterministic drift in \eqref{Langevin-multi-field}. Considering test scalar fields in de Sitter space for simplicity in the following, we thus take
\begin{eqnarray}
\frac{\partial \Psc}{\partial N}=\frac{1}{3 H_0^2} \nabla_I \left( V^{,I} \Psc \right)+\frac12 \left(  \frac{H_0}{2 \pi}\right)^2 \nabla_I \nabla^I \Psc
\label{FP-simple}
\end{eqnarray}
as our covariant equation that describes the corresponding diffusion of light scalar fields in a curved field space in the overdamped limit. Given that Eq.~\eqref{FP-simple} is the simplest multifield covariantisation of the corresponding single-field equation, one can wonder whether field space curvature invariants can also enter into this equation \cite{Vilenkin:1999kd}. To answer this question, one can use our knowledge of the relevant microphysics, \textit{i.e.} of quantum fluctuations of fields in NLSM. The Riemann curvature of the field space does enter at quadratic order in the action, however only in the effective mass matrix of the fluctuations. While in general this can have important consequences like the geometrical destabilisation of inflation \cite{Renaux-Petel:2015mga,Renaux-Petel:2017dia,Garcia-Saenz:2018ifx,Grocholski:2019mot}, we deduce that no curvature invariant should enter into the description of effectively \textit{light} fields, \textit{i.e.} with all the eigenvalues of the effective mass matrix much smaller than $H^2$, and that Eq.~\eqref{FP-simple} is thus sufficient for this purpose.

Let us now apply it to 2-field models with field space metric $(\partial \phi)^2+e^{2 b(\phi)} (\partial \psi)^2$. For a generic potential, one can easily derive from \eqref{FP-simple} evolution equations for arbitrary correlation functions, including for $n \geq 0$:
\begin{eqnarray}
\langle \phi^n \rangle'&=&-\frac{n}{3 H_0^2} \langle \phi^{n-1} V_{,\phi} \rangle \nonumber \\
&&\hspace{-1cm}+\frac12 \left( \frac{H_0}{2 \pi} \right)^2 \left[n(n-1) \langle \phi^{n-2} \rangle +n \langle \phi^{n-1} b_{,\phi} \rangle \right] \,.
\label{time-evolution-phin}
\end{eqnarray}
For sum separable potentials,
this system of equations has the interesting feature of involving only $\phi$. When $V_{,\phi}$ and the non-canonical function $b(\phi)$ are polynomials in $\phi$, it can thus be solved iteratively, starting from any initial distribution of $\phi$, and independently of the one of $\psi$. 
Let us concentrate on the interesting and simple case of an hyperbolic geometry, with $b(\phi)=- \phi/M$, and a quartic potential $\lambda \phi^4/4$. 
Formally Taylor-expanding the correlation functions as a function of the number of $e$-folds, $\langle \phi^n \rangle=\sum_{m=0}^{\infty} A^n_{\,m} N^m$, one obtains the recursive relations $(m+1) A^n_{\, m+1}=-\frac{n \lambda}{3 H_0^2} A^{n+2}_{\, m}+n(n-1) \frac{H_0^2}{8 \pi^2} A^{n-2}_{\, m}- \frac{n}{M} \frac{H_0^2}{8 \pi^2} A^{n-1}_{\, m}$, from which all the correlation functions can be deduced to arbitrary order $m$. With initial conditions $\langle \phi^n(0) \rangle =0$ for $n >0$, one finds for instance (see \cite{prep2} for more results)
\begin{eqnarray}
\langle \phi(N) \rangle= -\frac{H_0^2 N}{8 \pi^2 M} \left[1- \frac{\lambda N^2}{12 \pi^2}-\frac{\lambda H_0^2 N^3}{768 \pi^4 M^2} \nn
+\ldots \right]
\end{eqnarray}
\begin{eqnarray}
\hspace{-0.18cm}\langle \phi^2(N) \rangle= \frac{H_0^2 N}{4 \pi^2} \left[1+\frac{H_0^2 N}{16 \pi^2 M^2}-\frac{\lambda N^2}{6 \pi^2}-\frac{7 \lambda H_0^2 N^3}{384 \pi^4 M^2}+\ldots\right] \nn
\end{eqnarray}
Compared to the single-field case, the field space geometry breaks the $\mathrm{Z}_2$ symmetry of $\phi$, hence the appearance of non-zero correlations of odd powers of $\phi$. 
Note also 
the interplay between the effects of the potential and the field space curvature, coming with series expansion in $\lambda N^2/\pi^2$ and $H_0^2 N/(\pi^2 M^2)$ respectively.
This model, as a simple generalisation of the well studied $\lambda \phi^4$ single-field case, offers an interesting playground for studying stochastic effects in curved field space, and it would be interesting to confront these predictions to quantum field theory computations, which are comparably much more involved. 
Like in the single-field case, at late time, truncating the series to any finite order is misleading, but Eqs.~\eqref{FP-simple}--\eqref{time-evolution-phin} enable one to understand the asymptotic behaviour of the system. For a generic $b(\phi)$ and for sum-separable potentials, one can see from Eq.~\eqref{time-evolution-phin} that the statistical properties of $\phi$ can indeed be computed as in a single-field model with an effective potential $V_{{\rm eff}}(\phi)=V(\phi)-3 H_0^4/(8 \pi^2)\, b(\phi)$, with equilibrium distribution $P_{{\rm eq}} \propto e^{-8 \pi^2/(3 H_0^4)V_{{\rm eff}}(\phi)}$. Of course, this stationary distribution need not exist in general, \textit{i.e.} when it is not normalisable, for instance in the previous example with $\lambda=0$, which describes a simple Brownian motion around the linearly evolving shifted value $-H_0^2 N/(8 \pi^2 M)$. Eventually note that the consistency of our approach requires that its results should be such that all canonically normalised field fluctuations are light around the mean trajectory, which depends on the precise choice of potential $V(\phi,\psi)$ and the initial distribution of both fields. \\


{\bf Conclusions.}--- The stochastic approach aims at describing the long-wavelength part of quantum fields, in inflationary or de Sitter spacetimes, by a classical stochastic theory. As any effective theory, it comes with its regime of validity and its limitations, several of which that have already been emphasised in the literature. Here we highlighted an important conceptual issue of the stochastic approach that has not been previously pointed out, although similar difficulties are well identified in statistical physics (see \textit{e.g.}~\cite{vKampen-manifold,GRAHAM1985209,2016JSMTE..05.3207A,2017JPhA...50H5001C}): the \Ito interpretation of the Langevin equations does not respect general covariance under field redefinitions, while the Stratonovich one does, but at the expense of introducing spurious `frame' dependences. This feature holds in the overdamped description in terms of fields only (a description of the Einstein-Smoluchowski's type in the statistical physics' language), but also in a more complete phase space (Kramers) description. 
Despite these limitations, we used physical arguments to write down a manifestly covariant and physically motivated Fokker-Planck equation, with the aim of describing the quantum diffusion and the late-time behaviour of effectively light fields in a curved field space.
We applied it to test scalar fields in de Sitter space and showed that for a certain class of two-field models with derivative interactions, some statistical properties can be derived using a correspondence with a single-field model endowed with an effective potential.
We also studied a simple extension of the single-field $\lambda \phi^4$ theory to a hyperbolic field space geometry, making predictions that would be interesting to compare to first principles quantum field theory computations.

We stress that the difficulties we have described to formulate a generally covariant stochastic formalism seems to be the stochastic counterparts of the notoriously difficult problem of maintaining general covariance in quantum theories \cite{1957RvMP...29..377D,AlvarezGaume:1983ig,Grosche-87,Bastianelli:2006rx}. In particular, the way the Stratonovich prescription maintains formal general field space covariance at the expense of introducing local frame dependencies is reminiscent of the fact that the gravitational anomaly can be shifted from the general coordinate symmetry to the local Lorentz symmetry \cite{Bardeen:1984pm}. 
It would be very interesting to further investigate these links, and to determine whether the stochastic anomalies that we have described
can be circumvented by using techniques like the covariant background field expansion and the nonperturbative renormalisation group (see \textit{e.g.} \cite{Callan:1989nz,Guilleux:2015pma,Prokopec:2017vxx}), path-integral discretisation schemes that are more complex than simple $\alpha$-discretisations \cite{2018arXiv180609486C}, or by using the tools of open quantum systems and directly working with the quantum density matrix (see \textit{e.g.} \cite{Burgess:2014eoa,Burgess:2015ajz,Collins:2017haz}).


\textit{Acknowledgements.} We are grateful to Camille Aron, Cliff Burgess, Guillaume Faye, Richard Holman, Vivien Lecomte, Martin Lemoine, J\'er\^ome Martin, Patrick Peter, Cyril Pitrou, Gerasimos Rigopoulos, Julien Serreau, Alexei A. Starobinsky, Jean-Philippe Uzan and Vincent Vennin for useful discussions. S.RP is supported by the European Research Council (ERC) under the European Union's Horizon 2020 research and innovation programme (grant agreement No 758792, project GEODESI). Y. Tada was supported by grants from R\'egion \^Ile de France and Grand-in-Aid for JSPS Research Fellow (JP18J01992).


\appendix
\section{Stochastic observables in slow-roll single-field inflation}
\label{details-single-field}

The observable impact of stochastic effects on the curvature perturbation can be computed with the stochastic-$\delta N$ formalism~\cite{Fujita:2013cna,Fujita:2014tja} and its formulation in terms of the adjoint Fokker-Planck operator~\cite{Vennin:2015hra,Assadullahi:2016gkk,Pattison:2017mbe}. From a generic Fokker-Planck equation \eqref{FP}, one can deduce the statistical properties of the first passage time $\calN(\bm{X})$ from some point $\bm{X}$ to a given boundary. Indeed its characteristic (generating) function $\chi_\calN(\omega,\bm{X})=\braket{e^{i\omega\calN(\bm{X})}}$ can be shown to obey the adjoint FP 
equation~\cite{book-SDE}
\bae{
	\calL^\dagger_\FP(\bm{X})\chi_\calN(\omega,\bm{X})&=-i\omega\chi_\calN(\omega,\bm{X}) \nonumber \\
	\text{with} \quad \calL^\dagger_\FP(\bm{X})&=D^a\pdif{}{X^a}+\frac{1}{2}D^{ab}\frac{\partial^2}{\partial X^a\partial X^b}.
}
This can be readily applied to inflation, in which the $\delta N$ approach~\cite{Starobinsky:1986fxa} indicates that the observable curvature fluctuation coincides with the fluctuation of the number of $e$-folds elapsed until the system reaches the end-of-inflation surface. In particular, at leading order in slow-roll, its power spectrum $\calP_\zeta$ can be expressed in terms of the moments of $\calN$ as~\cite{Fujita:2013cna}
\begin{eqnarray}
\calP_\zeta(k)=\dif{\braket{\delta\calN^2}}{\braket{\calN}} |_{\langle\calN \rangle=\ln(k_\uf/k)}\,,
\end{eqnarray}
whereas the moments  $\calM_n=\braket{\calN^n}$
obey the recursive partial differential equations (PDE)
\bae{		
		\calL^\dagger_\FP\calM_n=-n\calM_{n-1}\,,
		\label{recursive}
}
from which the variance itself $\calC_2=\braket{\delta\calN^2}=\braket{\calN^2}-\braket{\calN}^2$ can be deduced to verify $\calL^\dagger_\FP\calC_2=-D^{ab}\pdif{\calM_1}{X^a}\pdif{\calM_1}{X^b}$\,. \\

When applied to the canonical slow-roll single-field description in terms of the Langevin equation \eqref{Langevin-single-field}, the equations \eqref{recursive} reduce to the simple set of ordinary differential equations
\bae{
	\bce{
		\dps
		v \calM_n^{\prime\prime}-\frac{v^\prime}{v}(1-\alpha v)\calM_n^\prime=-n\frac{\calM_{n-1}}{\Mpl^2}, \\
		\dps
		v \calC_2^{\prime\prime}-\frac{v^\prime}{v}(1-\alpha v)\calC_2^\prime=-2v{\calM_1^\prime}^2,
	}
}
where $v \equiv V/(24 \pi^2 \M^4)$ is the dimensionless potential in Planck units, $\alpha=0$ or $1/2$ in the \Ito or  Stratonovich interpretation respectively, and $'=d/d \phi$ here.
Its formal solution is given by
\bae{
	&\calM_n(\phi)= n\int^\phi_{\phi_\uf}\frac{\dd x}{\Mpl}\int^{\bar{\phi}_n}_x\frac{\dd y}{\Mpl} \frac{\calM_{n-1}(y)}{v^\alpha(x)v^{1-\alpha}(y)}\, {\cal K}(x,y) \label{eq: formal sol1} \\
	&\calC_2(\phi) =2\int^\phi_{\phi_\uf}\dd x\int^{\tilde{\phi}_2}_x\dd y\left(\frac{v(y)}{v(x)}\right)^\alpha{\calM_1^\prime}^2(y)\, {\cal K}(x,y),\label{eq: formal sol2}
}
with the kernel ${\cal K}(x,y)=\exp\left[\frac{1}{v(y)}-\frac{1}{v(x)}\right]$, $\phi_\uf$ represents the end of inflation and $\bar{\phi}_n$ and $\tilde{\phi}_2$ are constants of integration so that $v(y)>v(x)$. At leading order in slow-roll, where $\partial/\partial\log k \simeq -\partial\phi/\partial{\braket{\calN}}\times\partial/\partial\phi$, one thus obtains
\begin{equation}
\calP_\zeta=\calC_2^\prime/\calM_1^\prime\,, \quad \ns-1=-\frac{\calC_2^{\prime\prime}}{\calM_1^\prime\calC_2^\prime}+\frac{\calM_1^{\prime\prime}}{{\calM_1^\prime}^2}\,,
\label{observables}
\end{equation}
where the right-hand sides are evaluated at $\phi$, the mean inflaton value when the scale $k$ crosses the Hubble radius. \\

In the classical gravity regime where the inflationary energy scale is much smaller than the Planck scale, \textit{i.e.} $v(x)<v(y)\ll1$, the exponential suppression in $\exp\left[\frac{1}{v(y)}-\frac{1}{v(x)}\right]$ implies that, under certain conditions delineated below, the integrands in \eqref{eq: formal sol1}-\eqref{eq: formal sol2} acquire their main contributions from the region $y\sim x$. In this saddle-point limit, one obtains\footnote{These results consistently reduce to
those of Ref.~\cite{Vennin:2015hra} when $\alpha=0$, i.e. for the \Ito interpretation.}
\bae{
	\calM_1(\phi)&\simeq\int^\phi_{\phi_\uf}\frac{\dd x}{\Mpl^2}\frac{v}{v^\prime}(1+(1+\alpha)v-\eta_\cl), \label{eq: M1 approx} \\
	\calC_2(\phi)&\simeq2\int^\phi_{\phi_\uf}\frac{\dd x}{\Mpl^4}\frac{v^4}{{v^\prime}^3}(1+(6+3\alpha)v-5\eta_\cl),\\
	\calP_\zeta&\simeq\calP_{\zeta,\cl}(1+(5+2\alpha)v-4\eta_\cl),\\
	\ns-1& \simeq n_{{}_\text{S},\cl}-1\nonumber \\
	&+(-2(2+\alpha)\epsilon_V+3\eta_V)v-6\eta_V\eta_\cl+8\epsilon_V\xi_\cl,
}
where the expansion is reliable under the condition that the stochasticity parameters $\eta_\cl=v^2v^{\prime\prime}/{v^\prime}^2$ and $\xi_\cl=v^3v^{\prime\prime\prime}/{{v^\prime}^3}$ are small. In this regime, stochastic effects entail small corrections to the classical results $\calP_{\zeta,\cl}=\frac{2}{\Mpl^2}\frac{v^3}{{v^\prime}^2}$ and $n_{{}_\text{S},\cl}=1-6\epsilon_V-\eta_V$, where $\epsilon_V=\frac{\Mpl^2}{2}\frac{{v^\prime}^2}{v^2}$ and 
$\eta_V=\Mpl^2\frac{v^{\prime\prime}}{v}$ are the standard slow-roll parameters.

\begin{figure*}
	\centering
	\includegraphics[width=0.9\hsize]{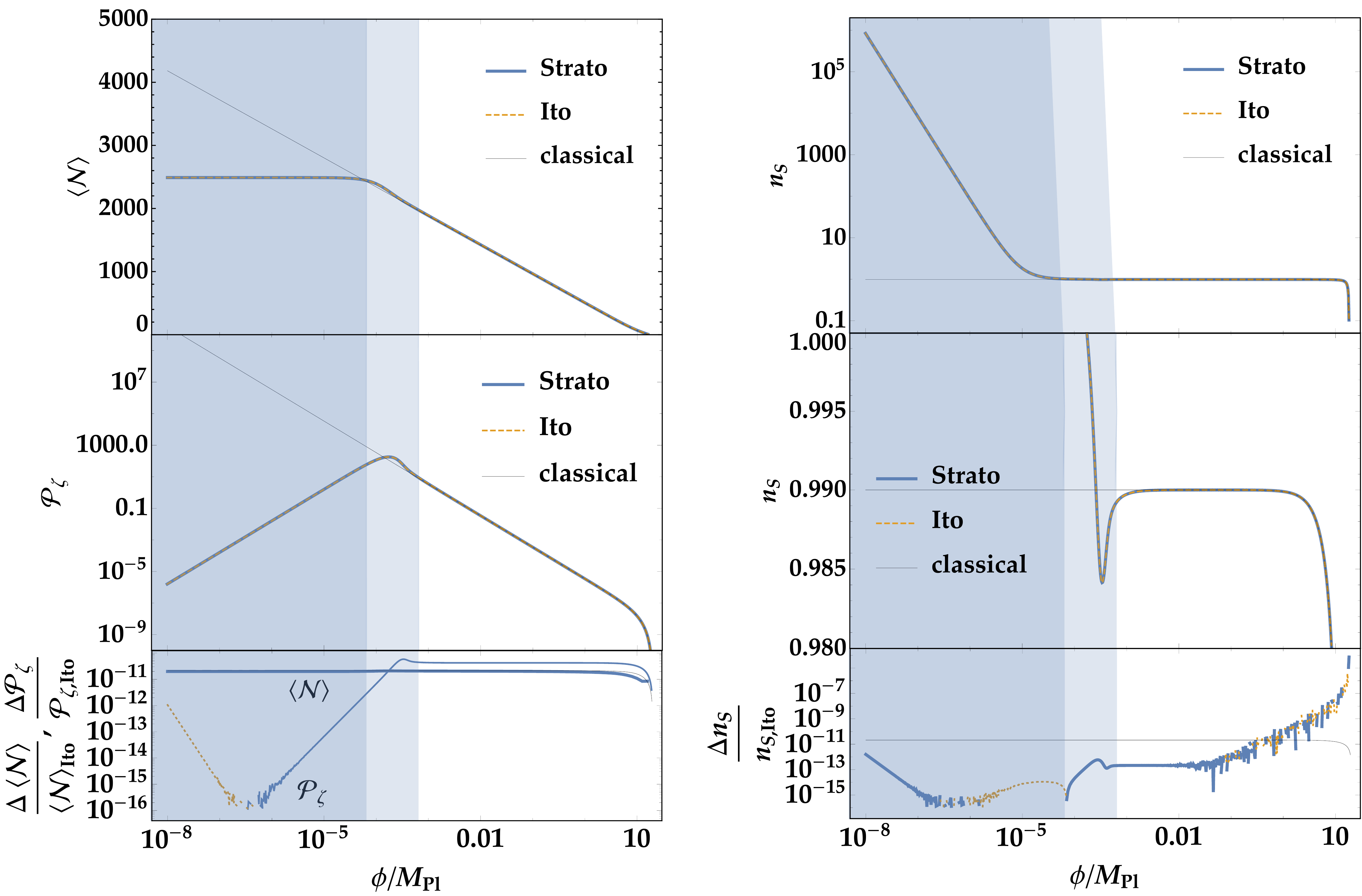}
	\caption{Numerical results for the mean number of $e$-folds $\braket{\calN}$ (top left), power spectrum $\calP_\zeta$ (bottom left), and spectral index $\ns$ (right), computed with the full stochastic formulae~\eqref{eq: formal sol1}-\eqref{observables} in the Stratonovich (``Strato") and It\^o (``Ito") interpretation, as well as the standard textbook ones (``classical"), in the hilltop model~(\ref{eq: hilltop}) with $\bar{\phi}_1=\tilde{\phi}_2=0$. Stochastic effects are non-negligible in the light blue region where $10^{-2} <|\eta_\cl| <1$ and very large in the dark blue region where $|\eta_\cl|>1$.
The lower plots enable one to see the very small fractional differences between the It\^o and Stratonovich results, which are expected to scale like $v(\phi)/2$, indicated by the thin black line. Blue plain lines represent positive values while orange dotted lines represent negative values.}
	\label{fig: hilltop}
\end{figure*}

Note that the $\alpha$-dependence of the above results, resulting from the choice between the \Ito or Stratonovich interpretation, is suppressed by $v\ll1$, and not by the stochasticity parameters $\eta_\cl$ and $\xi_\cl$. This suggests that this choice hardly affects observables, even when stochastic effects are large, \textit{i.e.} away from the saddle-point limit. We have checked this by resorting to numerical calculations for several choices of inflationary models. In Fig.~\ref{fig: hilltop} we present the results for the mean number of $e$-folds, the curvature power spectrum and the spectral index, for the hilltop model employed in Ref.~\cite{Vennin:2015hra} to exemplify stochastic effects:
\bae{\label{eq: hilltop}
	V(\phi)=\Lambda^4\left(1-\frac{\phi^2}{\mu^2}\right), \quad \Lambda=10^{-2}\Mpl, \quad \mu=20\Mpl\,,
}
with the same parameters $\bar{\phi}_1=\tilde{\phi}_2=0$ (one can check that observables away from the boundary conditions are insensitive to their precise choices). One can see there that the stochastic results substantially differ from the classical ones when the inflaton field traverses regions where $|\eta_\cl | \gtrsim 10^{-2}$, and even more so in regions with $|\eta_\cl | >1$ (note that $\xi_\cl=0$ for this potential).
More importantly, results derived from the \Ito or Stratonovich interpretations coincide to an excellent approximation, in the classical regime as well as when stochastic effects are large.

\section{Spurious frame-dependence in multifield situations}
\label{frame-dependence}

\begin{figure*}
	\centering
	\begin{tabular}{cc}
		\begin{minipage}{0.45\hsize}
			\centering
			\includegraphics[width=0.9\hsize]{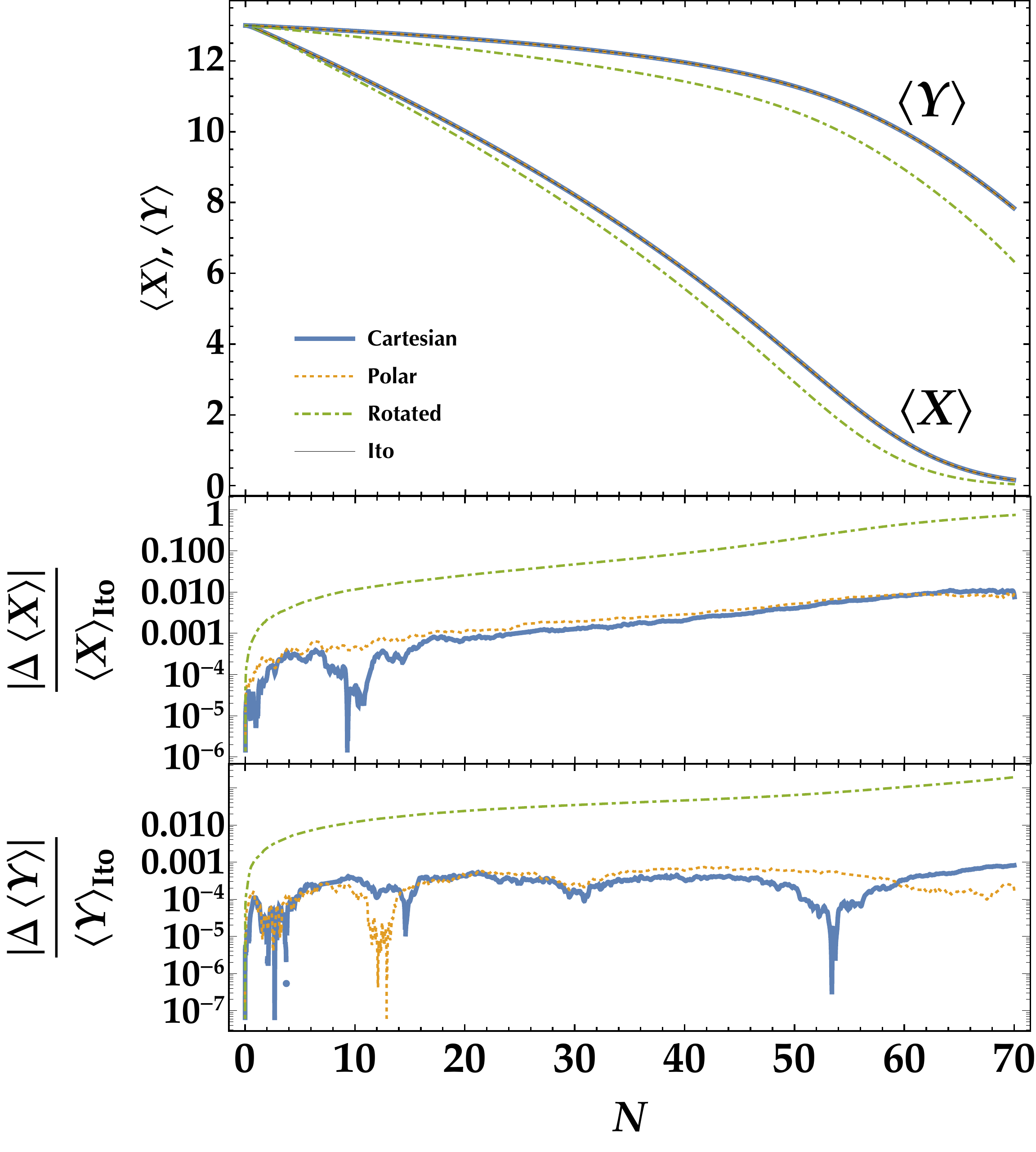}
		\end{minipage}
		\begin{minipage}{0.45\hsize}
			\centering
			\includegraphics[width=0.9\hsize]{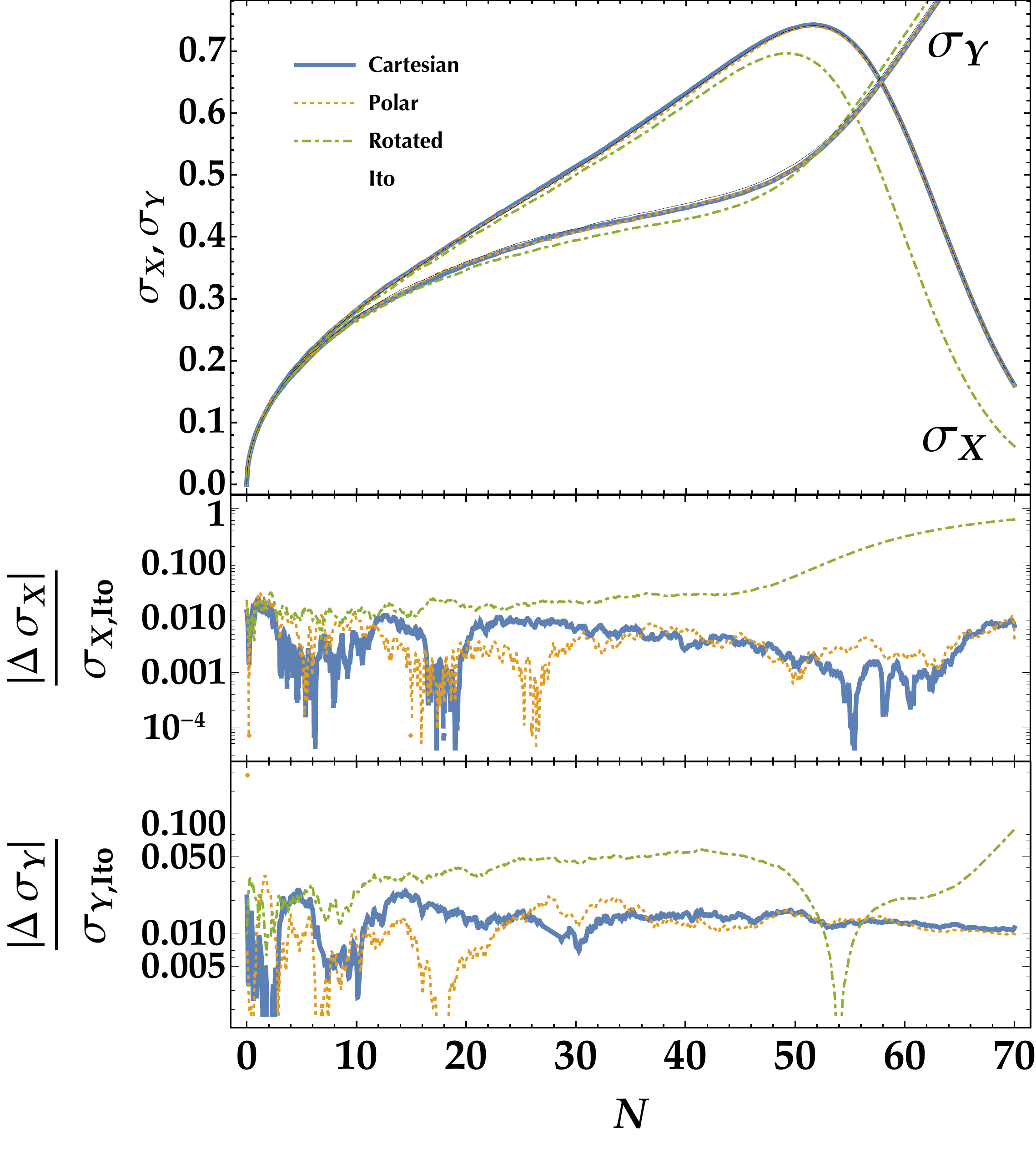}
		\end{minipage}
	\end{tabular}
	\caption{Time evolution of the averages (left) and standard deviations (right) of $X$ and $Y$ in the simple two-field model~(\ref{eq: double chaotic}) with initial conditions $X^{\rm ini}=Y^{{\rm ini}}=13 \M$. Averages are made out of $10^4$ realizations of the Langevin equations \eqref{Langevin-multi-field} with $\Xi^I=\frac{H}{2 \pi}e^I_\alpha \, \xi^\alpha$, and several choices of vielbeins $e^I_\alpha$: `Cartesian'~(\ref{Cartesian}), `Polar' and `Rotated' (respectively $m=1$ and $m=100$ in \eqref{eq: rotated noise}). We also included the result of the It\^o interpretation in the Cartesian coordinates. 
The lower plots enable one to visualise more easily the differences between the different curves, by displaying the fractional differences (in absolute value) with the It\^o result.}
	\label{fig: double_chaotic}
\end{figure*}

In this appendix, we show the concrete effects of different choices of vielbeins in the Stratonovich interpretation of the Langevin equations \eqref{Langevin-multi-field}. We consider the simple example of a two-field model with flat field space, whose Lagrangian reads 
\bae{\label{eq: double chaotic}
	\calL&=-\frac{1}{2}\partial_\mu X\partial^\mu X-\frac{1}{2}\partial_\mu Y\partial^\mu Y-\frac{1}{2}M_X^2X^2-\frac{1}{2}M_Y^2Y^2.
}
To make the effects of the noise easily visible, we choose unnaturally large masses $M_X=3M_Y=0.1\Mpl$, but the effects we discuss also apply to more realistic cases. We numerically solve $10^4$ realizations of the Langevin equations \eqref{Langevin-multi-field} with $\Xi^I=\frac{H}{2 \pi}e^I_\alpha \, \xi^\alpha$, in the Stratonovich interpretation, with initial conditions $X^{\rm ini}=Y^{{\rm ini}}=13 \M$, and for three different sets of vielbeins $e^I_\alpha$. We use the `Cartesian' fields $(X,Y)$ in the resolution but the results do not depend on this choice due to the general covariance of that interpretation. The first set of vielbeins that we choose is the natural ones in Cartesian coordinates:
\bae{
	(e^X_1=1,e^Y_1=0)\,, (e^X_2=0,e^Y_2=1) \,.
	\label{Cartesian}
}
The two other sets belong to the family of vielbeins defined by $(e^m)^{a}_\alpha = (\Omega^m)_\alpha^{\,\beta} e^{a}_\beta$, where the field-dependent rotation matrix is
\bae{\label{eq: rotated noise}
	 (\Omega^m)_\alpha^{\,\beta}=\bpme{
		\cos (m\Theta) & \sin(m\Theta) \\
		-\sin(m\Theta) & \cos(m\Theta)
	},
}
and where $\Theta=\tan^{-1}\frac{Y}{X}$ is the angle of the polar field space coordinates. We use $m=1$, which simply corresponds to the natural vielbeins used in polar coordinates (hence denoted as `Polar'), and $m=100$ (denoted as `Rotated'), in addition to the original Cartesian vielbeins \eqref{Cartesian} ($m=0$). The three corresponding results are shown in Fig.~\ref{fig: double_chaotic}, for the time-dependence of the average fields and of their standard deviations. One can see that, while the Cartesian and polar choices do not lead to appreciable differences for these variables, the choice $m=100$ significantly affects even the average dynamics. This is naturally expected from the corresponding FP equation \eqref{FP-Strato}, whose second-line contains a noise-induced drift $-\frac12  \left(\frac{H}{2 \pi}\right)^2 e^I_\alpha (\nabla_J e^J_\alpha)$, with $e^I_\alpha (\nabla_J e^J_\alpha)=m (X,Y)/(X^2+Y^2)$ in that case. For completeness, we also solved the corresponding Langevin equations with the \Ito interpretation in the Cartesian coordinates, whose results are almost indistinguishable from the Stratonovich results with Cartesian vielbeins.

\end{document}